\documentclass[letter,english,prl,twocolumn]{revtex4}
\usepackage{ae} 
\usepackage[T1]{fontenc}
\usepackage[ansinew]{inputenc}
\usepackage{amsmath}
\usepackage{amssymb}
\usepackage{amsfonts}
\usepackage[dvips]{graphicx}
\usepackage{ifthen}
\usepackage{eurosym}
\usepackage{babel}

\begin{document}

\title{Few-body reference data for multicomponent formalisms: Light nuclei molecules} 
\author{Ilkka Kyl\"anp\"a\"a$^{1}$, Tapio T.~Rantala$^{2}$ and David M.~Ceperley$^{1}$} 
\affiliation{$^{1}$Department of Physics, University of Illinois at Urbana-Champaign, Illinois 61801, USA\\$^{2}$Department of Physics, Tampere
  University of Technology, P.O.~Box 692, FI-33101 Tampere, Finland}
\date{\today}

\begin{abstract}
We present full quantum statistical energetics of some electron-light
nuclei systems. This is accomplished with the path integral Monte
Carlo method. The effects on energetics arising from the change in the
nuclear mass are studied.  The obtained results may serve as
reference data for the multicomponent density functional theory
calculations of light nuclei system. In addition, the results reported
here will enable better fitting of todays electron-nuclear energy
functionals, for which the description of light nuclei is most
challenging, in particular.
\end{abstract}

\maketitle


Density functional theory (DFT) is among the most succesful approaches
to calculate the electronic structure of atoms, molecules and
solids. A similar approach, however, including more degrees of freedom
was introduced in 2001 by Kreibich and Gross
\cite{PhysRevLett.86.2984}, and is called as multicomponent
density-functional theory (MCDFT).  In contrast to original form of
the DFT, MCDFT enables the complete quantum treatment of many particle
systems consisting of electrons and nuclei. As is well known, the
original form of DFT incorporates the Born--Oppenheimer approximation
for the nuclei \cite{PhysRev.136.B864,PhysRev.140.A1133}.

With the MCDFT approach it is possible to extend the success of DFT
into an entirely new field of applications, such as first-principles
calculation of electron-phonon coupling in solids
\cite{PhysRevB.69.115110,PhysRevB.69.199901}, polaronic motion
\cite{PhysRevB.69.075212} and positron scattering and annihilation
\cite{RevModPhys.82.2557,Walters05112010,Brawley05112010}. That is,
with MCDFT physical phenomena that depend on a strong coupling between
electronic and nuclear motion can be evaluated from first principles.

The original DFT is also known for its need of good functional forms,
especially for the exchange and correlation functional. One of the
most widely employed functional is the so-called local density
approximation (LDA), which uses the Monte Carlo data of the free
electron gas \cite{PhysRevLett.45.566} as a basic input. Proper
functional forms are also needed in the MCDFT scheme, for the
electron-nuclear energy functional
\cite{PhysRevA.78.022501,PhysRevLett.101.153001}, in particular. For
the present, the absence of good multicomponent reference data is
slowing down the development of new functional forms for the
MCDFT. The main difficulties are encountered in the description of
light nuclei.

In this brief report, we will provide few-body reference data for
light nuclei systems, which can be used in the development of better
MCDFT functionals and improving the present fits. This is accomplished
with full quantum statistical simulations using path integral Monte
Carlo (PIMC) approach \cite{Ceperley95}. The nuclear mass is given
values ranging from that of a positron to that of a proton described
by the following processes: $x^+e_2^-$, $x_2^+e^-$, $x_2^+e_2^-$ and
$x^+p^+e_2^-$, where $x^+$ goes from positron ($e^+$) to proton
($p^+$). A more detailed description of our approach is given in
Ref.~\cite{Kylanpaa11jcp}.


According to the Feynman formulation of the quantum statistical
mechanics \cite{Fey72} the partition function for interacting
distinguishable particles is given by the trace of the density matrix:
\begin{align}
Z
= \text{Tr}~\hat{\rho}(\beta)
= \int \rm{d} R_{0}\rm{d} R_{1} \ldots
\rm{d} R_{M-1} \prod_{i = 0}^{M-1}e^{-S(R_{i},R_{i+1};\tau)},\nonumber
\end{align}
where $\hat{\rho}(\beta) = e^{-\beta\hat{H}}$, $S$ is the action,
$\beta = 1/k_{\text{B}}T$, $\tau = \beta/M$, $R_{M}=R_{0}$ and $M$ is
called the Trotter number. In this paper, we use the pair
approximation in the action \cite{Storer68,Ceperley95} for the Coulomb
interaction of charges. Sampling in the configuration space is carried
out using the Metropolis procedure \cite{Metro53} with multilevel
bisection moves \cite{Chakravarty98}. The total energy is calculated
using the virial estimator \cite{Herman82}.

In the following we use atomic units, where the lengths, energies and
masses are given in units of the Bohr radius ($a_0$), hartree
($E_\text{h}$) and free electron mass ($m_e$), respectively. The
statistical standard error of the mean (SEM) with $2$SEM limits is
used as an error estimate for the observables.


In our model, all the particles are described as "boltzmannons",
i.e.~they obey the Boltzmann statistics. For the present study the
particles involved can be treated accurately as distinguishable
particles. This is possible by assigning spin-up to one electron and
spin-down to the other one, and applying the same for the positive
particles.  This is accurate enough, as long as the thermal energy is
well below that of the lowest electronic triplet excitation, $\Delta
E_{st}$. For the systems in consideration $\Delta E_{st}>0.18
E_{\rm{h}}$, the smallest being that of the Ps$_2$ molecule
\cite{Usukura02,PhysRevA.73.052712}. For more details on our model,
see Ref.~\cite{Kylanpaa11jcp}.

In the simulations we use $m_e = 1 = m_{e^+}$ as the mass of the
electrons and the positron, and for the protons we use $m_p =
1836.1527 m_e$. The simulations are carried out at $300$ K
temperature, and for the Trotter number we have chosen $M=8192$.  This
leads to ''time-step'' $\tau = \beta /M \approx 0.1285
E_{\rm{h}}^{-1}$, which ensures good enough accuracy in the case of
light nuclei --- the error is of order $\mathcal{O}(\tau^3)$. The
simulations apply the minimum image convention and a cubic simulation
cell, $V=(300a_0)^3$.

\begin{figure}[t]
\includegraphics{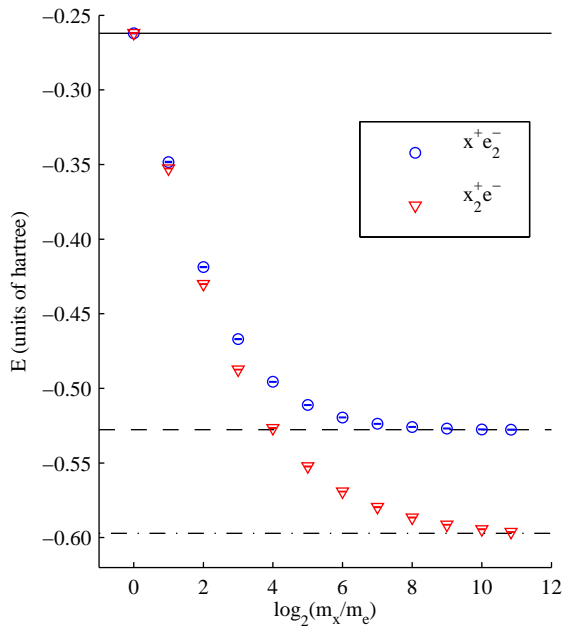}
\caption{\label{Fig1}Total energy as a function of mass of the nuclei,
  i.e.~positive particles. Blue circles show the energy for
  $x^+e_2^-$, and red down-triangles that for $x_2^+e^-$, where $x^+$
  goes from $e^+$ to $p^+$. The reference energies are given as solid
  line, dashed line and dash-dotted line corresponding to Ps$^-$ (as
  well as Ps$_2^+$), H$^-$ and H$_2^+$, respectively. }
\end{figure}

\begin{figure}[t]
\includegraphics{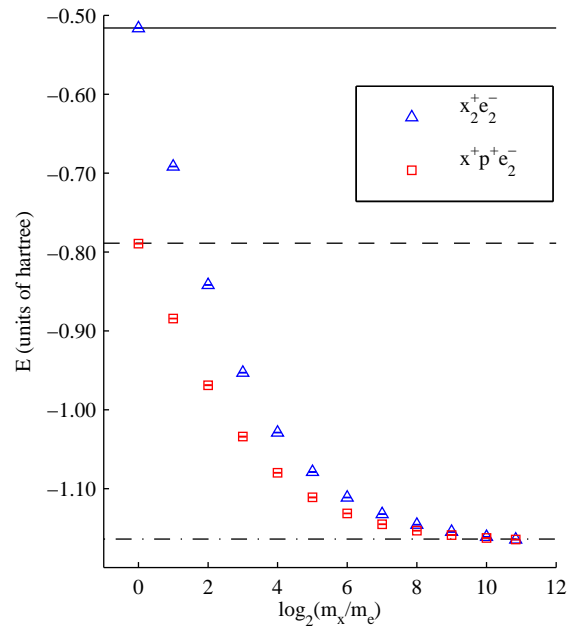}
\caption{\label{Fig2}Total energy as a function of mass of the nuclei,
  i.e.~positive particles. Blue up-triangles show the energy for
  $x_2^+e_2^-$, and red squares that for $x^+p^+e_2^-$, where $x^+$
  goes from $e^+$ to $p^+$. The reference energies are given as solid
  line, dashed line and dash-dotted line corresponding to Ps$_2$, PsH
  and H$_2$, respectively.}
\end{figure}

In Figs.~\ref{Fig1} and \ref{Fig2} we show the total energy as a
function of mass of the nuclei, i.e.~the positive particles: On the
left $x^+$ is equal to a positron, on the right $x^+$ corresponds to a
proton, and in the middle region we assign ten different masses for
the $x^+$ particle.

The total energies are also given in Table \ref{Table1}. The time-step
error affects mainly the fourth decimal in the total energies, which
can be validated by comparing the end-point values in Table
\ref{Table1} to high-accuracy zero Kelvin results. The comparison
shows that the difference between high accuracy results and our PIMC
values is less than $0.00094E_{\rm{h}}^{-1}$, which also confirms that
the order of the time-step error is $\mathcal{O}(\tau^3)$. Since the
fourth decimal is also uncertain due to statistical error estimate,
the present time-step error is considered acceptable. All energies
given in Table \ref{Table1} are from separate long enough simulations.
Due to the finite temperature present in our simulations there is a
small possibility for these molecules to dissociate even at the
temperature of $300$ K, however, none of our simulations experienced
dissociation.

\begin{table}[b]
\caption{\label{Table1}Total energies at different nuclear masses, see
  also Figs.~\ref{Fig1} and \ref{Fig2}. Energies are given in units of
  hartree with $2$SEM error estimates. Reading from up to down the
  $x^+$ in the table goes from positron to proton --- the mass of the
  particle increases.}
\begin{tabular}{ccccc}
\hline\hline
$\log_2(m_x/m_e)$ & $x^+e_2^-$ & $x_2^+e^-$ & $x_2^+e_2^-$ 
& $x^+p^+e_2^-$ \\
\hline
    0.0000  &  -0.2620(4) &   -0.2620(1) &   -0.5163(2)  &  -0.7895(3) \\
    1.0000  &  -0.3483(3) &   -0.3526(1) &   -0.6918(2)  &  -0.8845(3) \\
    2.0000  &  -0.4187(3) &   -0.4301(2) &   -0.8418(3)  &  -0.9690(4) \\
    3.0000  &  -0.4670(2) &   -0.4874(2) &   -0.9531(3)  &  -1.0341(3) \\
    4.0000  &  -0.4956(2) &   -0.5266(2) &   -1.0291(3)  &  -1.0800(3) \\
    5.0000  &  -0.5111(2) &   -0.5523(2) &   -1.0790(3)  &  -1.1112(3) \\
    6.0000  &  -0.5195(2) &   -0.5692(2) &   -1.1114(3)  &  -1.1315(3) \\
    7.0000  &  -0.5237(2) &   -0.5796(2) &   -1.1323(3)  &  -1.1451(3) \\
    8.0000  &  -0.5258(2) &   -0.5866(2) &   -1.1458(3)  &  -1.1533(3) \\
    9.0000  &  -0.5269(2) &   -0.5913(2) &   -1.1549(3)  &  -1.1588(3) \\ 
   10.0000  &  -0.5275(2) &   -0.5944(2) &   -1.1612(3)  &  -1.1627(3) \\
   10.8425  &  -0.5277(3) &   -0.5962(2) &   -1.1646(3)  &  -1.1647(3) \\
\hline\hline
\end{tabular}
\end{table}

The main difficulties in the MCDFT are related to the description of
light nuclei. Protons in small systems are already treated
reasonably. However, there definitely is room for improvement in that
case, and especially in case of positronic systems. The data presented
in Table \ref{Table1} will serve as a good reference data in the
development and fitting of electron-nuclear energy functionals. It
enables one to gradually go towards proper description of the lightest
and most difficult ''nucleus'', i.e.~the positron.

It should be pointed out, that proper density dependent reference data
will be essential for the success of MCDFT. Obtaining such results is
computationally demanding, however, the authors of this paper are
already working on it. For now, the results of this paper give useful
complementary information on the energetics of small light nuclei
systems, which can be used in the finding better fits for the
functionals.


We acknowledge CSC -- IT Center for Science Ltd.~and TCSC -- Tampere
Center for Scientific Computing for the allocation of computational
resources. For financial support we thank the Finnish Cultural
Foundation and the Physics Department of the University of Illinois at
Urbana-Champaign.

\end{document}